\documentclass[fp,twocolumn]{jpsj3}
\usepackage{txfonts}

\title{
  Lifshitz Transitions in Magnetic Phases of the Periodic Anderson Model
}

\author{Katsunori Kubo}
\inst{
Advanced Science Research Center, Japan Atomic Energy Agency,
Tokai, Ibaraki 319-1195, Japan}

\abst{
We investigate the reconstruction of a Fermi surface,
which is called a Lifshitz transition,
in magnetically ordered phases of the periodic Anderson model
on a square lattice
with a finite Coulomb interaction between $f$ electrons.
We apply the variational Monte Carlo method to the model
by using the Gutzwiller wavefunctions for the paramagnetic,
antiferromagnetic, ferromagnetic, and charge-density-wave states.
We find that an antiferromagnetic phase is realized around half-filling
and a ferromagnetic phase is realized
when the system is far away from half-filling.
In both magnetic phases, Lifshitz transitions take place.
By analyzing the electronic states,
we conclude that the Lifshitz transitions to large ordered-moment states
can be regarded as itinerant-localized transitions of the $f$ electrons.
}

\begin{document}
\maketitle

\section{Introduction}
The Fermi surface is an important ingredient
for characterizing a metallic state.
In general, the Fermi surface is affected by a phase transition
such as a magnetic transition.
On the other hand, the possibility of a phase transition
described by a change in the Fermi surface topology itself
has been proposed by Lifshitz~\cite{Lifshitz1960}.
In recent years, such Lifshitz transitions have been discussed
as a possible origin of some anomalies in heavy-fermion systems,
for example, the phase transition between ferromagnetic phases of UGe$_2$
under pressure~\cite{Oomi1998,Saxena2000,Tateiwa2001JPCM,Terashima2001,
  Tateiwa2001JPSJ,Settai2002,Terashima2002,Haga2002,Pfleiderer2002,Settai2003},
YbRh$_2$Si$_2$ under
a magnetic field~\cite{Tokiwa2005,Rourke2008,Pfau2013,Pourret2013,Naren2013},
and the transition between the antiferromagnetic phases
of CeRh$_{1-x}$Co$_x$In$_5$~\cite{Goh2008}.
Recently, Fermi surface reconstruction in the antiferromagnetic phase
of CeRhIn$_5$ under a magnetic field has also been reported~\cite{Jiao2015}.

Such a possibility of the existence of a Lifshitz transition
under a magnetic field and/or in a magnetically ordered state
in $f$-electron systems has been investigated theoretically for a long time.
%
Fermi surface reconstruction in an antiferromagnetic phase
is found in the Kondo lattice
model~\cite{Watanabe2007,Lanata2008,Martin2010,Hoshino2013}
and in the periodic Anderson model~\cite{Watanabe2009}.
%
Under a magnetic field or in a ferromagnetic phase,
there is a possibility of realizing a half-metallic state,
where only one spin band has a Fermi surface.
In the other phases, both spin bands have Fermi surfaces,
and thus a transition to the half-metallic state
from any of the other states inevitably accompanies a change
in the Fermi surface topology.
Indeed, such transitions to the half-metallic state have been found
in the Kondo lattice model~\cite{Irkhin1991,Watanabe2000,Kusminskiy2008,
  Beach2008,Peters2012PRL,Bercx2012,Peters2012PRB,Golez2013}
and in the periodic Anderson model~\cite{Reynolds1992,Dorin1993JAP,
  Dorin1993PRB,Kubo2013PSSC,Kubo2013PRB,Kubo2014,Wysokinski2014}.

In these theoretical studies,
while the models are similar,
the antiferromagnetic and ferromagnetic cases
are treated separately
except for a Kondo lattice model with the explicit inclusion of
antiferromagnetic and ferromagnetic Heisenberg interactions~\cite{Hoshino2013}.
For a further understanding of the Lifshitz transitions,
it is desirable to obtain a unified picture for both the magnetic cases.
In addition, in the above studies on the periodic Anderson model,
the Coulomb interaction $U$ between $f$ electrons is taken as
$U \rightarrow \infty$
except in the studies of the transition to the half-metallic state
by the slave-boson mean-field approximation~\cite{Reynolds1992,Dorin1993PRB}
and by a type of Gutzwiller approximation~\cite{Wysokinski2014}.
We also note that the Kondo lattice model is an effective model
of the periodic Anderson model in the limit of $U \rightarrow \infty$.
Thus, it is unclear whether the Lifshitz transition exists or not
even for a finite $U$ beyond these approximations.

In this work, we study the Lifshitz transitions
in the magnetic states of the periodic Anderson model with finite $U$
by applying the variational Monte Carlo method~\cite{Shiba1986,Watanabe2009}.
In this method, we do not introduce approximations
in evaluating physical quantities,
while we assume variational wavefunctions
as in the slave-boson mean-field and Gutzwiller approximation methods.
We investigate both the antiferromagnetic and ferromagnetic states
on an equal footing by varying the electron filling.
In particular, we analyze the physical quantities and energy gain
at the Lifshitz transitions to determine the characteristics of the transitions.
Preliminary results on the total energy for both magnetic cases
and the ordered moment in the antiferromagnetic case
have been reported in Ref.~\citen{Kubo2015}.

This paper is organized as follows.
In Sect.~\ref{model},
we explain the periodic Anderson model
and the variational wavefunctions in this study.
In Sect.~\ref{results},
we show the calculated results
for an antiferromagnetic case (Sect.~\ref{n1.917})
and for a ferromagnetic case (Sect.~\ref{n1.5}).
We calculate the energy and physical quantities
such as the ordered moment and effective mass.
We also discuss the nature of the phase transitions in the magnetic phases
with the aid of analyses
of the energy components and the momentum distribution functions.
Then, we discuss the Fermi surface structures in the magnetic phases.
The last section is devoted to a summary.

\section{Model and Method}\label{model}
The periodic Anderson model is given by
\begin{equation}
  \begin{split}
    \mathcal{H}=&\sum_{\mib{k} \sigma}\epsilon_{\mib{k}}
    c^{\dagger}_{\mib{k} \sigma}c_{\mib{k} \sigma}
    +\sum_{i \sigma}\epsilon_f n_{f i \sigma}\\
    -&V\sum_{\mib{k} \sigma}(f^{\dagger}_{\mib{k} \sigma}c_{\mib{k} \sigma}
                            +c^{\dagger}_{\mib{k} \sigma}f_{\mib{k} \sigma})
    +U\sum_{i}n_{f i \uparrow}n_{f i \downarrow},
  \end{split}
\end{equation}
where
$c^{\dagger}_{\mib{k} \sigma}$
and
$f^{\dagger}_{\mib{k} \sigma}$
are the creation operators of the conduction and $f$ electrons,
respectively, with momentum $\mib{k}$ and spin $\sigma$.
$n_{f i \sigma}$ is the number operator
of the $f$ electron with spin $\sigma$ at site $i$.
$\epsilon_{\mib{k}}$ is the kinetic energy of the conduction electron,
$\epsilon_f$ is the $f$-electron level,
$V$ is the hybridization matrix element,
and $U$ is the onsite Coulomb interaction between $f$ electrons.
Here, we consider only the nearest-neighbor hopping for the conduction electrons
on a square lattice,
and the kinetic energy is given by
$\epsilon_{\mib{k}}=-2t(\cos k_x + \cos k_y)$,
where $t$ is the hopping integral and we set the lattice constant as unity.

We apply the variational Monte Carlo method
to the model~\cite{Shiba1986,Watanabe2009}.
As the variational wavefunction,
we consider the following Gutzwiller wavefunction:
\begin{equation}
  | \psi \rangle
  =P | \phi \rangle,
\end{equation}
where
\begin{equation}
  P=\prod_{i}[1-(1-g)n_{f i \uparrow}n_{f i \downarrow}]
\end{equation}
is a projection operator with the variational parameter $g$.
This parameter controls the probability of the double occupancy
of the $f$ electrons on the same site.
In the limiting cases,
$g=1$, i.e., $P=1$ for $U=0$
and $g=0$, i.e., the double occupancy is prohibited for $U \rightarrow \infty$.
For a finite $U$ as in this study,
we have to determine $g$ between zero and unity to minimize the energy.
$| \phi \rangle$ is the one-electron part of the wavefunction.
In the present study, we choose the one-electron part as the ground state
of a mean-field-type effective Hamiltonian.

For the paramagnetic or ferromagnetic state, i.e., for a uniform state,
we consider the following effective Hamiltonian:
\begin{equation}
  H_{\text{eff}}=\sum_{\mib{k} \sigma}
  (c^{\dagger}_{\mib{k} \sigma} \ f^{\dagger}_{\mib{k} \sigma})
  \begin{pmatrix}
    \epsilon_{\mib{k}} & -\tilde{V}_{\sigma} \\[1.5ex]
    -\tilde{V}_{\sigma} & \tilde{\epsilon}_{f \sigma}
  \end{pmatrix}
  \begin{pmatrix}
    c_{\mib{k} \sigma} \\[1.5ex]
    f_{\mib{k} \sigma}
  \end{pmatrix},
\end{equation}
where $\tilde{V}_{\sigma}$ is the effective hybridization matrix element
and $\tilde{\epsilon}_{f \sigma}$ is the effective $f$-level.
They are variational parameters.
For the paramagnetic state, they do not depend on spin $\sigma$.

For the antiferromagnetic state,
we consider a state with the ordering vector $\mib{Q}=(\pi,\pi)$.
Then, the effective Hamiltonian is given by
\begin{equation}
  \begin{split}
    H_{\text{eff}}&=\sum_{\mib{k} \sigma}
    (c^{\dagger}_{\mib{k} \sigma} \ f^{\dagger}_{\mib{k} \sigma} \
    c^{\dagger}_{\mib{k}+\mib{Q} \sigma} \
    f^{\dagger}_{\mib{k}+\mib{Q} \sigma})\\
    \times&
    \begin{pmatrix}
      \epsilon_{\mib{k}} & -\tilde{V} &
      \sigma \tilde{\epsilon}_{c \mib{Q}} &
      -\sigma \tilde{V}_{\mib{Q}} \\[1.5ex]
      -\tilde{V} & \tilde{\epsilon}_{f} &
      -\sigma \tilde{V}_{\mib{Q}} &
      \sigma \tilde{\epsilon}_{f \mib{Q}} \\[1.5ex]
      \sigma \tilde{\epsilon}_{c \mib{Q}} &
      -\sigma \tilde{V}_{\mib{Q}} &
      \epsilon_{\mib{k}+\mib{Q}} & -\tilde{V} \\[1.5ex]
      -\sigma \tilde{V}_{\mib{Q}} &
      \sigma \tilde{\epsilon}_{f \mib{Q}} &
      -\tilde{V} & \tilde{\epsilon}_{f}
    \end{pmatrix}
    \begin{pmatrix}
      c_{\mib{k} \sigma} \\[1.5ex]
      f_{\mib{k} \sigma} \\[1.5ex]
      c_{\mib{k}+\mib{Q} \sigma} \\[1.5ex]
      f_{\mib{k}+\mib{Q} \sigma}
    \end{pmatrix},
  \end{split}
\end{equation}
where $\mib{k}$-summation runs over the folded Brillouin zone
of the antiferromagnetic state
and $\sigma$ in front of the parameters stands for $+$ ($-$)
for the up-spin (down-spin) states.
The parameters with a tilde are variational parameters.
$\tilde{\epsilon}_{c \mib{Q}}$ and $\tilde{\epsilon}_{f \mib{Q}}$
play roles similar to mean fields.
In addition, we consider $\tilde{V}_{\mib{Q}}$,
which describes the staggered component
of the effective hybridization matrix element in the antiferromagnetic state.

For the charge-density-wave state with $\mib{Q}=(\pi,\pi)$,
we can also consider a similar effective Hamiltonian,
but we find that the charge-density-wave state does not become
the ground state within the parameters that we have investigated.
In the Kondo lattice model,
the possibility of the charge-density-wave state
has been discussed.~\cite{Hirsch1984, Otsuki2009,Peters2013,Misawa2013}
To discuss this possibility in the periodic Anderson model,
we need to investigate a much wider parameter space,
e.g., by varying $U$,
since the charge-density-wave state is considered to be realized
in an intermediate coupling regime in the Kondo lattice model.
Thus, we show results only for the paramagnetic, ferromagnetic,
and antiferromagnetic states in the following.

To construct $|\phi \rangle$,
we fix the number of electrons per site of each spin $\sigma$, $n_{\sigma}$.
In the paramagnetic and antiferromagnetic states, $n_{\uparrow}=n_{\downarrow}$.
For the ferromagnetic state,
the magnetization $M=n_{\uparrow}-n_{\downarrow}$ is a parameter
characterizing the state.

For each state, we evaluate the energy by the Monte Carlo method,
and optimize the variational parameters that minimize the energy.
Then, we compare the energies of these states
with the same electron density $n=n_{\uparrow}+n_{\downarrow}$
and determine the ground state.
Other physical quantities can also be calculated by the Monte Carlo method
with the optimized variational parameters.

In this study, we set $U=8t$ and $V=t$,
that is, $U$ is the same as the bandwidth of the conduction electrons
and $V$ is much smaller than the bandwidth.
The calculations are carried out for an $L \times L$ lattice with $L=12$.
The boundary condition is antiperiodic for the $x$-direction
and periodic for the $y$-direction.

\section{Results}\label{results}

\subsection{Around half-filling: $n=1.917$}\label{n1.917}
First, we show the results around half-filling ($n=2$).
We set the number of electrons per site $n$ to $276/12^2 = 1.917$.

\begin{figure}
  \includegraphics[width=0.99\linewidth]
  {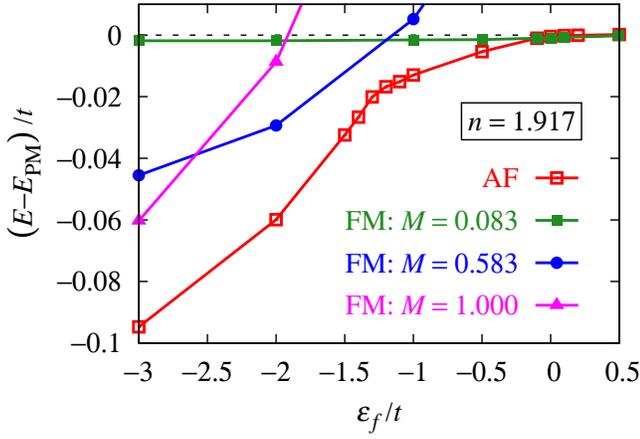}
  \caption{\label{E_n1.917}
    (Color online)
    Energy as a function of $\epsilon_f$
    measured from that of the paramagnetic state $E_{\text{PM}}$
    for the antiferromagnetic (AF) state (open squares)
    and for the ferromagnetic (FM) states
    with $M=0.083$ (solid squares),
         $M=0.583$ (circles),
    and  $M=1$     (triangles).
    $U/t=8$, $V/t=1$, and $n=1.917$.
  }
\end{figure}

Figure~\ref{E_n1.917} shows the energy $E$ per site
of the antiferromagnetic (AF)
and ferromagnetic (FM) states measured from
that in the paramagnetic (PM) state $E_{\text{PM}}$
as a function of $\epsilon_f$.
For the ferromagnetic states, 
we show the results for $M=0.083$, 0.583, and 1.
The state with $M=0.083$ is the half-metallic state
for this filling, i.e., $M=n_{\uparrow}-n_{\downarrow}=1-(n-1)=2-n$.

In a wide parameter region,
we find that the antiferromagnetic state is the ground state.
At $\epsilon_f/t \gtrsim -0.1$,
the half-metallic state with $M=0.083$ has the lowest energy,
while the difference in energy is not visible on this scale.
The energy gain of this weak ferromagnetic state is very small,
and it may become unstable against the paramagnetic state
when we improve the variational wavefunction.
Thus, we simply ignore this ferromagnetic state here
and concentrate on the antiferromagnetic state.
In the antiferromagnetic state,
there is a bend in the energy at $\epsilon_f/t \simeq -1.3$.
The discontinuity in the first derivative of the energy indicates
a first-order phase transition.

\begin{figure}
  \includegraphics[width=0.99\linewidth]
  {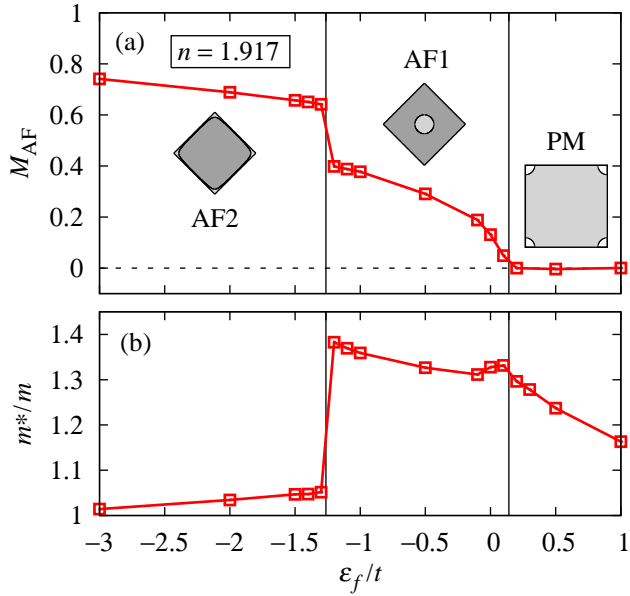}
  \caption{\label{mAF_inv_Dnk_n1.917}
    (Color online)
    (a) Antiferromagnetic moment $M_{\text{AF}}$ and
    (b) effective mass
    as functions of $\epsilon_f$
    for $U/t=8$, $V/t=1$, and $n=1.917$.
    The vertical lines denote the phase boundaries.
    We draw the Fermi surface in each phase in (a):
    only the lower hybridized band is occupied in the lightly shaded areas
    and both the hybridized bands are occupied in the darkly shaded areas.
  }
\end{figure}

In Fig.~\ref{mAF_inv_Dnk_n1.917}(a), we show the antiferromagnetic moment
as a function of $\epsilon_f$.
The antiferromagnetic moment is defined as
\begin{equation}
  M_{\text{AF}}=
  \frac{1}{N}\sum_{i}
  e^{i \mib{Q} \cdot \mib{r}_i}
  \langle n_{i \uparrow}-n_{i \downarrow} \rangle,
\end{equation}
where $N=L^2$ is the number of lattice sites,
$\mib{r}_i$ is the position of site $i$,
$n_{i \sigma}$ is the number operator of the electrons
with spin $\sigma$ at site $i$,
and $\langle \cdots \rangle$ denotes the expectation value.
By decreasing $\epsilon_f$,
$M_{\text{AF}}$ develops from zero around $\epsilon_f \simeq 0.1$.
This seems to be a continuous phase transition,
although we cannot discriminate it from a weak first-order transition
in the present numerical calculation.
By decreasing $\epsilon_f$ further,
we find a jump in $M_{\text{AF}}$ at $\epsilon_f/t \simeq -1.3$.
This is a first-order phase transition as is already recognized
from the energy (Fig.~\ref{E_n1.917}).

Here, we call the antiferromagnetic phase with smaller $M_{\text{AF}}$
($\epsilon_f/t \gtrsim -1.3$) AF1
and that with larger $M_{\text{AF}}$
($\epsilon_f/t \lesssim -1.3$) AF2.
For each phase, we can draw the Fermi surface
by using the obtained variational parameters in the one-electron part
[see insets in Fig.~\ref{mAF_inv_Dnk_n1.917}(a)].
We will discuss these Fermi surface structures later.

\begin{figure}
  \includegraphics[width=0.99\linewidth]
  {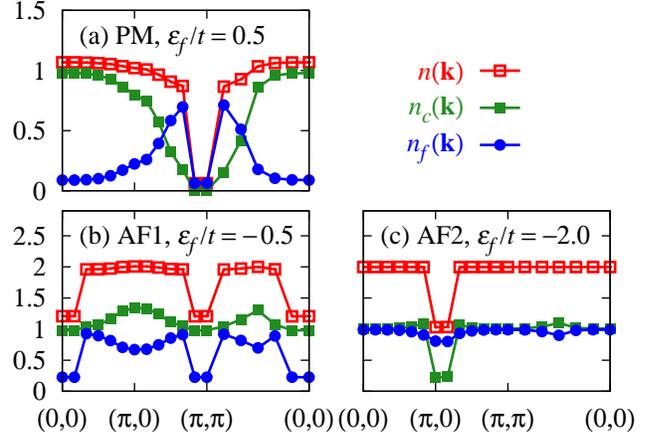}
  \caption{\label{nk_n1.917}
    (Color online)
    Momentum distribution functions
    $n(\mib{k})$ (open squares),
    $n_c(\mib{k})$ (solid squares), and
    $n_f(\mib{k})$ (circles)
    for
    (a) $\epsilon_f/t=0.5$ (PM),
    (b) $\epsilon_f/t=-0.5$ (AF1), and
    (c) $\epsilon_f/t=-2$ (AF2).
    $U/t=8$, $V/t=1$, and $n=1.917$.
    Owing to the antiperiodic boundary condition for the $x$-direction,
    we shift $k_x$ by $\pi/L$, e.g., $(\pi,\pi)$ in the figures
    actually means the point $(\pi-\pi/L,\pi)$.
  }
\end{figure}

In Fig.~\ref{mAF_inv_Dnk_n1.917}(b),
we show the effective mass $m^*$
defined by the jump $\Delta n(\mib{k}_{\text{F}})$
in the momentum distribution function $n(\mib{k})$ (see Fig.~\ref{nk_n1.917})
at the Fermi momentum $\mib{k}_{\text{F}}$:
\begin{equation}
  \frac{m^*}{m}=\frac{1}{\Delta n(\mib{k}_{\text{F}})},
\end{equation}
where $m$ is the bare mass.
Here, $\Delta n(\mib{k}_{\text{F}})$ is defined as the jump
in the momentum distribution function
along $(\pi,0)$--$(\pi,\pi)$ for the paramagnetic state
and
along $(  0,0)$--$(\pi,  0)$ for the antiferromagnetic states.
In the paramagnetic state,
$m^*$ increases as $\epsilon_f$ decreases,
since the number of $f$ electrons increases
and correlation effects become stronger.
At the PM-AF1 phase transition,
$m^*$ does not change significantly since it is a continuous transition.
In the AF1 state,
$m^*$ continues to increase except around the PM-AF1 phase boundary.
On the other hand, in the AF2 state, the effective mass becomes lighter
since the magnetic moment develops sufficiently
and the correlation effects become weak.

\begin{figure}
  \includegraphics[width=0.99\linewidth]
  {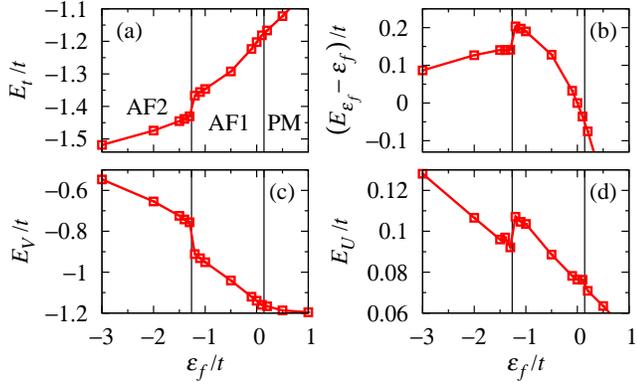}
  \caption{\label{E_decomposition_n1.917}
    (Color online)
    Components of energy as functions of $\epsilon_f$:
    (a) kinetic energy of the conduction electrons,
    (b) site energy of the $f$ electrons $E_{\epsilon_f}=\epsilon_f n_f$
        measured from $\epsilon_f$,
    (c) energy of the hybridization, and
    (d) energy of the Coulomb interaction.
    $U/t=8$, $V/t=1$, and $n=1.917$.
  }
\end{figure}

Note that the symmetry is the same between the AF1 and AF2 states.
To determine what characterizes the AF1-AF2 transition,
we decompose the energy into four terms:
the kinetic energy of the conduction electrons,
\begin{equation}
  E_t=\frac{1}{N}\sum_{\mib{k} \sigma} \epsilon_{\mib{k}}
  \langle c^{\dagger}_{\mib{k} \sigma}c_{\mib{k} \sigma} \rangle,
\end{equation}
the site energy of the $f$ electrons,
\begin{equation}
  E_{\epsilon_f}=\epsilon_f n_f,
\end{equation}
the hybridization energy,
\begin{equation}
  E_V=-\frac{V}{N}\sum_{\mib{k} \sigma}
  \langle f^{\dagger}_{\mib{k} \sigma}c_{\mib{k} \sigma}
         +c^{\dagger}_{\mib{k} \sigma}f_{\mib{k} \sigma} \rangle,
\end{equation}
and the Coulomb interaction,
\begin{equation}
  E_U=\frac{U}{N}\sum_{i} \langle n_{f i \uparrow}n_{f i \downarrow} \rangle,
\end{equation}
where $n_f$ is the expectation value of the number of $f$ electrons
per site.
Figure~\ref{E_decomposition_n1.917} shows the decomposed terms
as functions of $\epsilon_f$.
At the PM-AF1 transition, these terms change smoothly.
At the transition from AF1 to AF2,
the gain in the hybridization $E_V$ decreases,
while the gains in $E_t$ and $E_{\epsilon_f}$ increase.
This indicates that the conduction and $f$ electrons are relatively decoupled
in the AF2 state.
The change in $E_U$ at the AF1-AF2 transition is small in comparison with
the other terms.

\begin{figure}
  \includegraphics[width=0.99\linewidth]
  {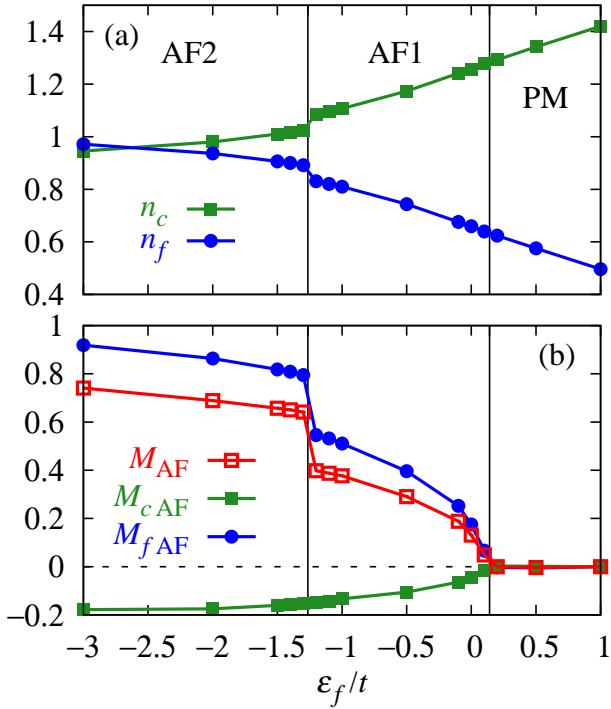}
  \caption{\label{n_mAF_n1.917}
    (Color online)
    Contributions of the conduction and $f$ electrons
    to the electron number and to the antiferromagnetic moment
    as functions of $\epsilon_f$.
    (a) Numbers of conduction electrons, $n_c$ (squares),
    and $f$ electrons, $n_f$ (circles), per site.
    (b) Total antiferromagnetic moment, $M_{\text{AF}}$ (open squares),
    the antiferromagnetic moment of the conduction electrons,
    $M_{c \text{AF}}$ (solid squares), and
    the antiferromagnetic moment of the $f$ electrons,
    $M_{f \text{AF}}$ (circles).
    $U/t=8$, $V/t=1$, and $n=1.917$.
  }
\end{figure}

In Fig.~\ref{n_mAF_n1.917}(a),
we show the $\epsilon_f$ dependences of the occupancies of
the conduction and $f$ electrons.
$n_c$ is the expectation value of the number
of conduction electrons per site.
The $f$-electron number $n_f$ increases as the $f$ level decreases,
and in the AF2 state, it almost reaches unity.
In Fig.~\ref{n_mAF_n1.917}(b),
we show the antiferromagnetic moments
of the conduction electrons, $M_{c \text{AF}}$,
and of the $f$ electrons, $M_{f \text{AF}}$,
as functions of $\epsilon_f$.
They are defined as
\begin{align}
  M_{c \text{AF}}&=\frac{1}{N}\sum_{i} e^{i \mib{Q} \cdot \mib{r}_i}
  \langle n_{c i \uparrow}-n_{c i \downarrow} \rangle,\\
  M_{f \text{AF}}&=\frac{1}{N}\sum_{i} e^{i \mib{Q} \cdot \mib{r}_i}
  \langle n_{f i \uparrow}-n_{f i \downarrow} \rangle,
\end{align}
where $n_{c i \sigma}$ is the number operator of the conduction electron
at site $i$ with spin $\sigma$.
In the periodic Anderson model,
the conduction and $f$ electrons tend to have spins that are opposite
to each other at the same site.
Thus, $M_{c \text{AF}}$ and $M_{f \text{AF}}$ have opposite signs.
The total magnetic moment $M_{\text{AF}}$ is mainly composed
of the $f$ component.
In the AF2 state, $M_{f \text{AF}}$ is near to unity.
This also indicates that the $f$ electrons are almost localized
in the AF2 state.

The momentum distribution functions in each phase
are shown in Fig.~\ref{nk_n1.917}.
For the paramagnetic or ferromagnetic state,
the momentum distribution functions are defined as
\begin{align}
  n_{c \sigma}(\mib{k})&=
  \langle c^{\dagger}_{\mib{k} \sigma}c_{\mib{k} \sigma} \rangle,\\
  n_{f \sigma}(\mib{k})&=
  \langle f^{\dagger}_{\mib{k} \sigma}f_{\mib{k} \sigma} \rangle,\\
  n_{\sigma}(\mib{k})&=n_{c \sigma}(\mib{k})+n_{f \sigma}(\mib{k}).
\end{align}
For the antiferromagnetic state:
\begin{align}
  n_{c \sigma}(\mib{k})&=
  \langle
   c^{\dagger}_{\mib{k}            \sigma}c_{\mib{k}            \sigma}
  +c^{\dagger}_{\mib{k}+\mib{Q} \sigma}c_{\mib{k}+\mib{Q} \sigma}
  \rangle,\\
  n_{f \sigma}(\mib{k})&=
  \langle
   f^{\dagger}_{\mib{k}            \sigma}f_{\mib{k}            \sigma}
  +f^{\dagger}_{\mib{k}+\mib{Q} \sigma}f_{\mib{k}+\mib{Q} \sigma}
  \rangle,\\
  n_{\sigma}(\mib{k})&=n_{c \sigma}(\mib{k})+n_{f \sigma}(\mib{k}).
\end{align}
They do not depend on the spin $\sigma$ in the paramagnetic
and antiferromagnetic states:
$n_{c \uparrow}(\mib{k})=n_{c \downarrow}(\mib{k})=n_{c}(\mib{k})$,
$n_{f \uparrow}(\mib{k})=n_{f \downarrow}(\mib{k})=n_{f}(\mib{k})$, and
$n_{  \uparrow}(\mib{k})=n_{  \downarrow}(\mib{k})=n    (\mib{k})$.

\begin{figure}
  \includegraphics[width=0.99\linewidth]
  {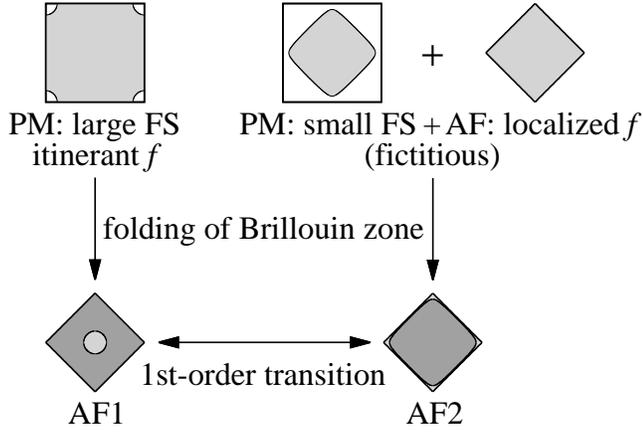}
  \caption{\label{FS_n1.917}
    Fermi surface (FS) structure in each phase
    obtained for $U/t=8$, $V/t=1$, and $n=1.917$.
    Only the lower hybridized band is occupied in the lightly shaded areas
    and both the hybridized bands are occupied in the darkly shaded areas.
    We also draw the Fermi surface structure
    for a fictitious localized $f$-electron state
    with a small Fermi surface composed only of the conduction electrons.
  }
\end{figure}

In Fig.~\ref{nk_n1.917},
we recognize most of the Fermi momenta on the symmetry axes
by the clear jumps in $n(\mib{k})$
even in the finite-size lattice in the present study.
While $n(\mib{k})$ should also have jumps around $(\pi/2,\pi/2)$
in the AF2 state (see Fig.~\ref{FS_n1.917}),
we could not detect them in the lattice with the present size.
In the PM and AF1 states,
the jumps in the total momentum distribution function $n(\mib{k})$
are mainly composed of the $f$ contribution $n_f(\mib{k})$.
On the other hand, in the AF2 state,
the jumps are mainly due to the conduction-electron contribution
$n_c(\mib{k})$.
In the AF2 state, $n_f(\mib{k})$ is almost flat, that is,
the $f$ electrons are nearly localized in the real space.

In Fig.~\ref{FS_n1.917}, we show the Fermi surface structure
in each state.
In the paramagnetic state,
there is a small hole pocket around ($\pi,\pi$)
since it is near half-filling.
In the present theory, the paramagnetic state is always regarded
as an itinerant $f$ state, that is,
the $f$-electron state contributes to the volume of the Fermi surface.
In the AF1 state, we obtain a hole pocket centered at (0,0).
This Fermi surface can be obtained by simply folding the paramagnetic
Fermi surface.
Thus, the AF1 state is naturally connected to the paramagnetic state,
and in this sense, it is regarded as an itinerant $f$ state.
In the AF2 state, the Fermi surface is different from that in the AF1 state.
Thus, we can discriminate these antiferromagnetic states
on the basis of the Fermi surface structures,
while the symmetries of these states are the same.

The Fermi surface in the AF2 state can be obtained by considering
a fictitious small Fermi surface state.
If each site has one perfectly localized $f$ electron
decoupled from the conduction electrons
and these $f$ electrons order antiferromagnetically,
then we obtain a small Fermi surface composed only of the conduction electrons
with filling $n-1$.
By combining these conduction- and $f$-electron states,
we obtain the same Fermi surface as in the AF2 state.
This means that the AF2 state can be interpreted as a localized $f$ state.
Note, however,
that the conduction and $f$ electrons are not completely decoupled.

The AF1-AF2 transition is of first order,
since the AF2 Fermi surface cannot be obtained by continuously deforming
that in the AF1 state.
The PM-AF1 transition can become of first order in general,
while in the present calculation, it is continuous.

In CeRh$_{1-x}$Co$_x$In$_5$,
there are two antiferromagnetic phases as in the present theory.
The change in the Fermi surface between the antiferromagnetic phases
is observed by the de Haas-van Alphen measurement~\cite{Goh2008}.
The variation in the effective mass
deduced from the de Haas-van Alphen measurement
as a function of $x$ is similar to
that shown in Fig.~\ref{mAF_inv_Dnk_n1.917}(b).
While the transition between the antiferromagnetic phases
in CeRh$_{1-x}$Co$_x$In$_5$
is a commensurate-incommensurate transition,
the present theory should have some relevance to this material,
for example, the mechanism of the change in the effective mass.

\subsection{Far away from half-filling: $n=1.5$}\label{n1.5}
\begin{figure}
  \includegraphics[width=0.99\linewidth]
  {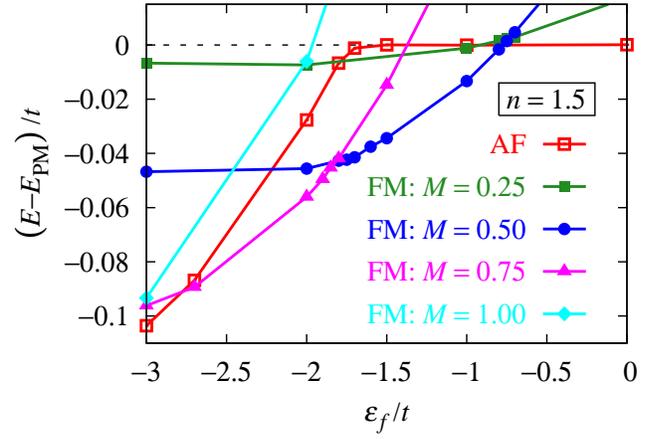}
  \caption{\label{E_n1.5}
    (Color online)
    Energy as functions of $\epsilon_f$
    measured from that of the paramagnetic state $E_{\text{PM}}$
    for the antiferromagnetic state (open squares)
    and the ferromagnetic states
    with $M=0.25$ (solid squares),
         $M=0.5$  (circles),
         $M=0.75$ (triangles),
    and  $M=1$    (diamonds).
    $U/t=8$, $V/t=1$, and $n=1.5$.
  }
\end{figure}

Next, we show the results for $n=1.5$.
We expect that the antiferromagnetism becomes weak
for such a case far away from half-filling
and there is a chance of stabilizing a ferromagnetic state.
Figure~\ref{E_n1.5} shows the energy as functions of $\epsilon_f$ for $n=1.5$.
In contrast to the case around half-filling,
the ferromagnetic state has a lower energy than the antiferromagnetic state
in a wide parameter region.
We note that while the antiferromagnetic state has the lowest energy
at $\epsilon_f/t=-3$ in Fig.~\ref{E_n1.5},
ferromagnetic states with $M \simeq 0.9$ (not shown) have lower energy there.

\begin{figure}
  \includegraphics[width=0.99\linewidth]
  {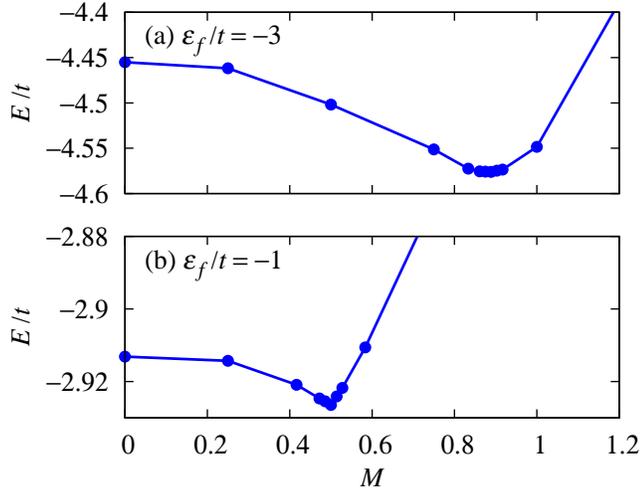}
  \caption{\label{E_FM_n1.5_efm1_and_efm3}
    (Color online)
    Energy as a function of magnetization $M$:
    (a) $\epsilon_f/t=-3$ and (b) $\epsilon_f/t=-1$.
    $U/t=8$, $V/t=1$, and $n=1.5$.
  }
\end{figure}

To determine the magnetization $M$ for each $\epsilon_f$,
we calculate the energy as a function of $M$.
In Fig.~\ref{E_FM_n1.5_efm1_and_efm3},
we show results for $\epsilon_f/t=-3$ and $-1$ as examples.
For $\epsilon_f/t=-3$ [Fig.~\ref{E_FM_n1.5_efm1_and_efm3}(a)],
the energy becomes minimum at $M \simeq 0.89$.
For $\epsilon_f/t=-1$ [Fig.~\ref{E_FM_n1.5_efm1_and_efm3}(b)],
the energy becomes minimum at $M=0.5$.
The state with $M=0.5$ is the half-metallic state for this filling.
We find a cusp in the energy at the minimum point $M=0.5$.
This indicates a gap in the spin excitation for the half-metallic state,
since the magnetic susceptibility $\chi$ is given by
$\frac{d^2 E(M)}{dM^2}=\chi^{-1}$
and a cusp in $E(M)$ results in $\chi=0$.
This gap originates from the hybridization gap between the up-spin bands.

\begin{figure}
  \includegraphics[width=0.99\linewidth]
  {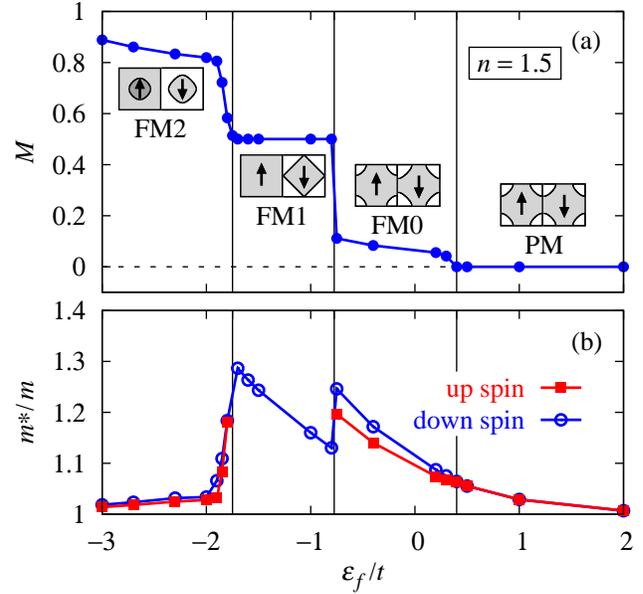}
  \caption{\label{m_inv_Dnk_n1.5}
    (Color online)
    (a) Magnetization $M$ and
    (b) effective mass for up-spin (squares) and down-spin (circles) states
    as functions of $\epsilon_f$
    for $U/t=8$, $V/t=1$, and $n=1.5$.
    The vertical lines denote the phase boundaries.
    In the FM1 state, the Fermi surface is absent for the up-spin state
    and we cannot define the effective mass for it.
    We draw the Fermi surface in each phase in (a):
    only the lower hybridized band is occupied in the lightly shaded areas
    and both the hybridized bands are occupied in the darkly shaded area.
  }
\end{figure}

In Fig.~\ref{m_inv_Dnk_n1.5}(a),
we show the magnetization as a function of $\epsilon_f$.
By decreasing $\epsilon_f$, $M$ gradually develops from zero
around $\epsilon_f \simeq 0.4$.
For $-1.7 \lesssim \epsilon_f \lesssim -0.8$,
we obtain the half-metallic state, $M=0.5$.
The magnetization is flat in this region.
By decreasing $\epsilon_f$ further,
$M$ increases again and asymptotically reaches unity.
In the following, we call
the low-magnetization state ($M<0.5$) FM0,
the half-metallic state ($M=0.5$) FM1, and
the high-magnetization state ($M>0.5$) FM2.
These ferromagnetic states have the same symmetry,
but we can discriminate them on the basis of the Fermi surface structures
as shown in Fig.~\ref{m_inv_Dnk_n1.5}(a).
We will discuss the details of the Fermi surface structures later.

In Fig.~\ref{m_inv_Dnk_n1.5}(b),
we show the $\epsilon_f$ dependence of the effective mass for each spin state.
Here, the effective mass is defined
along $(\pi,0)$--$(\pi,\pi)$ for the PM, FM0, and FM1 phases
and along $(0,0)$--$(\pi,0)$ for the FM2 phase.
Note that in the half-metallic phase FM1,
there is no Fermi surface for the up-spin state
and we cannot define the effective mass for the up-spin electrons.
In the PM, FM0, and FM1 states,
the effective mass increases as $\epsilon_f$ decreases
except at the FM0-FM1 boundary.
In the FM2 state, $m^*$ decreases as $\epsilon_f$ decreases
since the ordered moment becomes large.

\begin{figure}
  \includegraphics[width=0.99\linewidth]
  {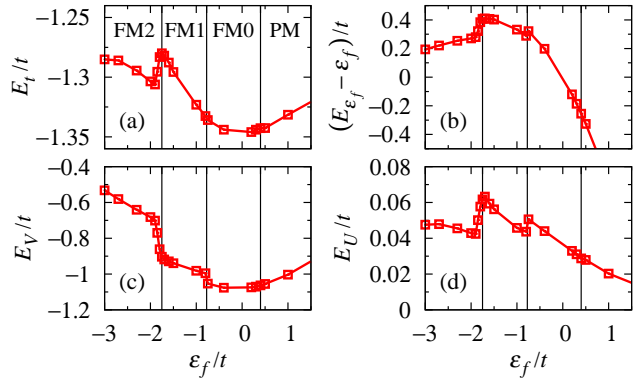}
  \caption{\label{E_decomposition_n1.5}
    (Color online)
    Components of energy as functions of $\epsilon_f$:
    (a) kinetic energy of the conduction electrons,
    (b) site energy of the $f$ electrons $E_{\epsilon_f}=\epsilon_f n_f$
        measured from $\epsilon_f$,
    (c) energy of the hybridization, and
    (d) energy of the Coulomb interaction.
    $U/t=8$, $V/t=1$, and $n=1.5$.
  }
\end{figure}

In Fig.~\ref{E_decomposition_n1.5}, we show the components of the energy.
The changes in these components at the phase boundaries are weak
except for the FM1-FM2 transition.
At the transition from FM1 to FM2,
the gain in the hybridization energy $E_V$ is reduced,
while the gains in the kinetic energy $E_t$ of the conduction electrons
and the site energy $E_{\epsilon_f}$ of the $f$ electrons increase.
This indicates that the conduction and $f$ electrons are nearly decoupled
in the FM2 phase, as in the AF2 phase of $n=1.917$.
The change in $E_U$ at the FM1-FM2 transition is smaller
than those in the other terms.

\begin{figure}
  \includegraphics[width=0.99\linewidth]
  {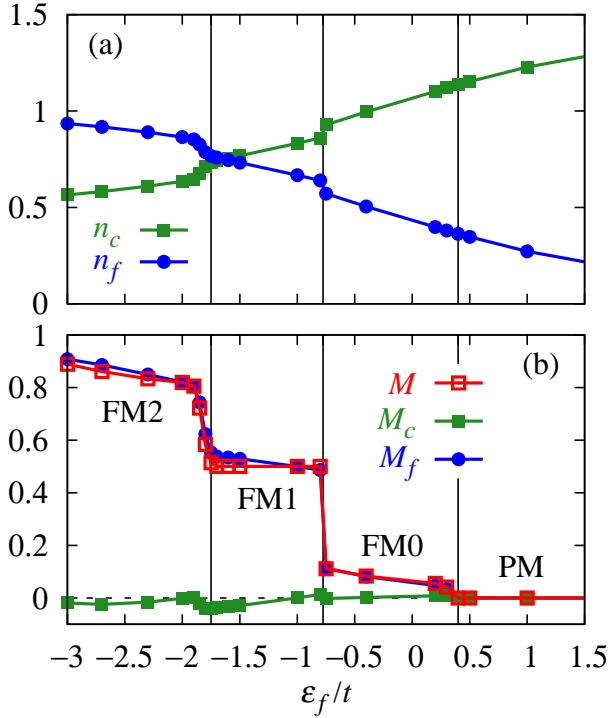}
  \caption{\label{n_m_n1.5}
    (Color online)
    Contributions of the conduction and $f$ electrons
    to the electron number and to the magnetization
    as functions of $\epsilon_f$.
    (a) Numbers of conduction electrons, $n_c$ (squares),
    and $f$ electrons, $n_f$ (circles), per site.
    (b) Total magnetization $M$ (open squares),
    the magnetization of the conduction electrons,
    $M_c$ (solid squares), and
    the magnetization of the $f$ electrons,
    $M_f$ (circles).
    $U/t=8$, $V/t=1$, and $n=1.5$.
  }
\end{figure}

In Fig.~\ref{n_m_n1.5}(a),
we show the $\epsilon_f$ dependences of $n_c$ and $n_f$.
$n_f$ increases as $\epsilon_f$ decreases
and reaches almost unity in the FM2 phase.
In Fig.~\ref{n_m_n1.5}(b),
we show the magnetization of the conduction and $f$ electrons,
$M_c$ and $M_f$, respectively,
and the total magnetization $M$.
$M_c$ and $M_f$ are given by
\begin{align}
  M_c&=\frac{1}{N}\sum_{i}
  \langle n_{c i \uparrow}-n_{c i \downarrow} \rangle,\\
  M_f&=\frac{1}{N}\sum_{i}
  \langle n_{f i \uparrow}-n_{f i \downarrow} \rangle.
\end{align}
At most data points, $M_c$ and $M_f$ have opposite signs.
Although at some points, $M_c$ and $M_f$ have the same sign,
the absolute values of $M_c$ are very small there.
The $f$-electron contribution $M_f$ dominates the total magnetization $M$,
and $M_f$ is nearly unity in the AF2 phase.

In actual situations, we should take different values of the $g$-factors
for the conduction and $f$ electrons.
Thus, the total magnetization is not proportional to $M=M_c+M_f$.
However, $M_c$ is small and the overall features
in the total magnetization will not change, e.g.,
the magnetization will remain almost flat in the FM1 phase.

\begin{figure}
  \includegraphics[width=0.99\linewidth]
  {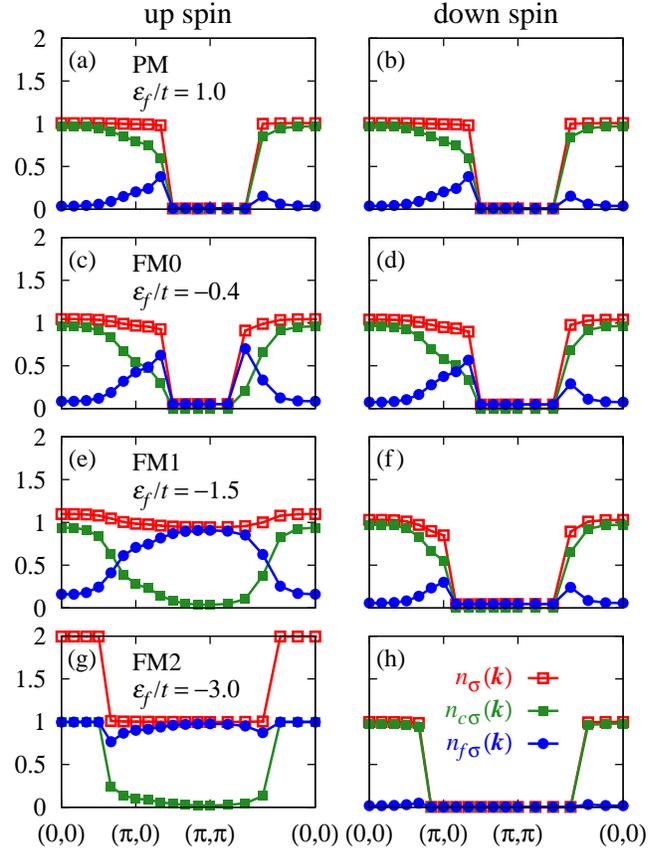}
  \caption{\label{nk_n1.5}
    (Color online)
    Momentum distribution functions
    $n_{\sigma}(\mib{k})$ (open squares),
    $n_{c \sigma}(\mib{k})$ (solid squares), and
    $n_{f \sigma}(\mib{k})$ (circles)
    for
    (a),(b) $\epsilon_f/t=1$ (PM),
    (c),(d) $\epsilon_f/t=-0.4$ (FM0),
    (e),(f) $\epsilon_f/t=-1.5$ (FM1), and
    (g),(h) $\epsilon_f/t=-3$ (FM2).
    The left (right) panels show those of up-spin (down-spin) states.
    $U/t=8$, $V/t=1$, and $n=1.5$.
    Owing to the antiperiodic boundary condition for the $x$-direction,
    we shift $k_x$ by $\pi/L$, e.g., $(\pi,\pi)$ in the figures
    actually means $(\pi-\pi/L,\pi)$.
  }
\end{figure}

Figure~\ref{nk_n1.5} shows the momentum distribution functions in each phase.
In the PM phase, they do not depend on spin.
In the FM0 phase, the number of up-spin electrons increases
and the hole Fermi surface around $(\pi,\pi)$ shrinks for the up-spin state.
For the down-spin state, the hole Fermi surface should become larger,
but, owing to the small magnetization
and the small lattice size in the present study,
we cannot detect the change.
The number of $f$ electrons is larger than that in the PM state,
and the contribution of the $f$ electrons increases,
particularly around the Fermi momenta.
In the FM1 phase, the Fermi surface for the up-spin state disappears
as is recognized from the absence of jumps in $n_{\uparrow}(\mib{k})$.
In the FM2 phase, the jumps in $n_{\sigma}(\mib{k})$ at the Fermi momenta
are mainly composed of $n_{c \sigma}(\mib{k})$.
$n_{f \sigma}(\mib{k})$ is nearly flat
and the $f$ electrons are almost localized in the real space.

\begin{figure}
  \includegraphics[width=0.99\linewidth]
  {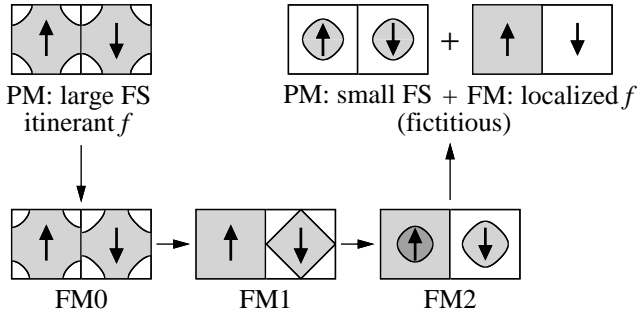}
  \caption{\label{FS_n1.5}
    Fermi surface (FS) structure in each phase
    obtained for $U/t=8$, $V/t=1$, and $n=1.5$.
    Only the lower hybridized band is occupied in the lightly shaded areas
    and both the hybridized bands are occupied in the darkly shaded area.
    The Fermi surface for FM0 is obtained at $\epsilon_f/t=-0.75$
    and that for FM2 is obtained at $\epsilon_f/t=-3$.
    In the other phases, the Fermi surface does not change with $\epsilon_f$.
    We also draw the Fermi surface structure
    for a fictitious localized $f$-electron state
    with a small Fermi surface composed of the conduction electrons.
  }
\end{figure}

In Fig.~\ref{FS_n1.5}, we show the Fermi surface structure
in each state.
The Fermi surface in the PM state is what is called a large Fermi surface
with the $f$-electron contribution.
In the FM0 phase, the hole Fermi surface of the up-spin state shrinks,
and in the FM1 phase, it disappears.
In the FM2 phase, the up-spin electrons partially occupy the upper band,
and as a result, the Fermi surface structures for up- and down-spin states
become similar to each other.

This Fermi surface in the FM2 state
can be understood from a localized $f$ picture.
The Fermi surface in the FM2 state is approximately decomposed into
a fictitious localized $f$ state with complete polarization
and a paramagnetic small Fermi surface of the conduction electrons
with filling $n_c=n-1$.
Thus, the FM2 state is regarded as a localized $f$ state.

In the present calculation,
the FM0-FM1 transition is of first order
and the other transitions are continuous.
However, in general,
it is possible for each of them to occur through either
a first-order transition or a continuous transition,
since each Fermi surface can be continuously deformed into the others.

In UGe$_2$ under pressure,
there are two ferromagnetic phases,
which probably correspond to FM1 and FM2 in this study.
We are uncertain whether the FM0 phase can be eliminated
by tuning the parameters.
If we ignore the weak magnetization in the FM0 state, the overall behaviors
of the magnetization and the effective mass in Fig.~\ref{m_inv_Dnk_n1.5}
are similar to those as functions of pressure
in UGe$_2$~\cite{Oomi1998,Saxena2000,Tateiwa2001JPCM,Settai2002,Pfleiderer2002}.
In addition, the Fermi surface reconstructions at the phase transitions
are also observed
in the de Haas-van Alphen measurements~\cite{Terashima2001,Settai2002,
  Terashima2002,Haga2002,Settai2003}.
Thus, we expect that the FM1-FM2 transition in UGe$_2$
is a Lifshitz transition corresponding to the present theory.

\section{Summary}
By applying the variational Monte Carlo method,
we have investigated both the antiferromagnetic and ferromagnetic states
of the periodic Anderson model with finite $U$ on an equal footing.
We have found the antiferromagnetic states (AF1 and AF2)
around half-filling ($n=1.917$)
and the ferromagnetic states (FM0, FM1, and FM2)
for a case far away from half-filling ($n=1.5$).

The weak magnetic states, AF1 and FM0, are naturally connected
to the paramagnetic state with a large Fermi surface.
On the other hand, the large ordered-moment states, AF2 and FM2,
can be regarded as localized $f$ states with a small Fermi surface.
This has been confirmed from the behavior of several quantities:
the effective mass, the energy components such as the hybridization energy,
the momentum distribution functions,
and the Fermi surface structure.
We have also found a half-metallic state FM1 between FM0 and FM2 for $n=1.5$.

These magnetic phases are characterized by the Fermi surface structure,
and the transitions between them are Lifshitz transitions
without symmetry breaking.
This is consistent with the previous studies with $U \rightarrow \infty$
that separately discussed the antiferromagnetic and ferromagnetic cases.
While we have not found a feature peculiar to a finite-$U$ case,
it gives justification for the use
of $U \rightarrow \infty$ in related theories.

In the present study,
by carefully analyzing several quantities,
we have reached a unified picture of the Lifshitz transitions
for both the antiferromagnetic and ferromagnetic cases.
In particular, we have clearly shown that
both the transitions to the large ordered-moment states, AF2 and FM2, are
itinerant-localized transitions of the $f$ electrons.

However, in the present theory,
we could not obtain a large effective mass,
since the large ordered-moment states appear
before the effective mass is enhanced substantially.
To attain a coherent understanding
of the heavy-fermion state and its magnetic order,
we need further breakthroughs,
such as improving the wavefunction and/or revising the model.
These are important future problems.

\section*{Acknowledgments}
The author thanks Y. Tokunaga for useful comments,
particularly on the energy decomposition.  
This work was supported by JSPS KAKENHI Grant Numbers 23740282 and 15K05191.



\end{document}